\documentclass{article}
\pdfoutput=1
\usepackage{GEM_workshop_2024}

\usepackage{amsmath,amsfonts,bm}

\def\eqref#1{equation~\ref{#1}}

\def\1{\bm{1}}

\DeclareMathAlphabet{\mathsfit}{\encodingdefault}{\sfdefault}{m}{sl}
\SetMathAlphabet{\mathsfit}{bold}{\encodingdefault}{\sfdefault}{bx}{n}

\usepackage{hyperref}
\usepackage{url}
\usepackage{algorithm}
\usepackage{graphicx}
\usepackage{algpseudocode}
\usepackage{float}
\usepackage{tabularx}
\usepackage{multirow}
\usepackage{caption}
\usepackage{subcaption}

\title{GeoDirDock: Guiding Docking Along Geodesic Paths}

\author{Raúl Miñán, Javier Gallardo \thanks{Work done during an internship at Nostrum Biodiscovery.}, Álvaro Ciudad, Alexis Molina \thanks{Author has an affiliation with Barcelona Supercomputing Center.}\\
Department of Artificial Intelligence\\
Nostrum Biodiscovery S.L.\\
Barcelona, Spain\\
\texttt{\{alvaro.ciudad,alexis.molina\}@nostrumbiodiscovery.com}
}

\iclrfinalcopy 
\begin{document}

\maketitle
\begin{abstract} 
This work introduces GeoDirDock (GDD), a novel approach to molecular docking that enhances the accuracy and physical plausibility of ligand docking predictions. GDD guides the denoising process of a diffusion model along geodesic paths within multiple spaces representing translational, rotational, and torsional degrees of freedom. Our method leverages expert knowledge to direct the generative modeling process, specifically targeting desired protein-ligand interaction regions. We demonstrate that GDD significantly outperforms existing blind docking methods in terms of RMSD accuracy and physicochemical pose realism. Our results indicate that incorporating domain expertise into the diffusion process leads to more biologically relevant docking predictions. Additionally, we explore the potential of GDD for lead optimization in drug discovery through angle transfer in maximal common substructure (MCS) docking, showcasing its capability to predict ligand orientations for chemically similar compounds accurately.
\end{abstract}

\section{Introduction}
In drug discovery, molecular docking is pivotal for probing interactions between molecules and specific protein cavities, also known as binding pockets \citep{Meng2011-at}. These regions on protein surfaces are targets for docking algorithms, aiming to establish the most stable configurations of ligands, or ligand \textit{poses}, for assessing potential drug-receptor interactions. Docking traditionally relies on empirical data from \textit{in vitro} experiments, such as crystal structures and molecular dynamics simulations, to locate these cavities for molecular binding assessment. Renowned methods like Glide, Vina, and rDock leverage physics-based scoring and a rich database of molecular interactions to navigate this process \citep{glide,vina,rdock}.

While traditional methods have long been a cornerstone in computational chemistry, the field experienced a paradigm shift with DiffDock \citep{Corso2022DiffDockDS}. This seminal contribution postulated molecular docking as a generative task, marking a significant shift through the development of a diffusion method that operates, without a prior over the binding site location, across the various degrees of freedom inherent in molecular docking. However, limitations appear during prospective applications, due to its blind docking strategy. Directed docking, which employs expert knowledge or pocket prediction to identify binding regions, has demonstrated superior performance compared to blind docking \citep{blind-gz}.
Additionally, the use of ligand root mean squared distance (RMSD) as the sole performance metric has been contested for not fully capturing ligand position accuracy, raising questions about the effectiveness of recent deep learning innovations in this field. \citep{posebysters, dodeeplearning}.

In this work, we introduce GeoDirDock, a diffusion method that integrates expert knowledge into the diffusion process over translations, rotations, and torsions, to direct the diffusion process towards a desired region of the protein structure. Our results show that this directed docking approach significantly improves upon blind docking methods, achieving close-to-ground truth RMSD conformations in self-docking scenarios and demonstrating enhanced performance in both RMSD accuracy and physicochemical pose plausibility. Moreover, we validate the robustness of our approach through a maximal common substructure docking test, highlighting its effective generalization across diverse and previously unseen ligand chemistries.

\section{Related work}

\textbf{Traditional Prior-Informed Docking Methods}. Traditional docking evaluates how likely molecules are to bind to specific areas of proteins, mainly focusing on binding pockets. These methods capitalize on data from crystal structures, literature, and molecular dynamics simulations to set pose search parameters and find the most suitable docking positions, using physics-based scoring methods to measure molecular interactions \citep{Meng2011-at,smina,gnina}. However, the effectiveness of these methods largely relies on having access to high-quality data, which can be a limitation in cases where such data is scarce.

\textbf{Diffusion-Based Molecular Modeling}. Recent advancements in diffusion models have relied upon the concept of blind docking, which does not rely on pre-existing receptor-ligand binding information. DiffDock, for example, replaces traditional methods with a denoising process that manipulates translations, rotations, and torsion angles in a $\mathbb{T}^3 \times SO(3) \times SO(2)^m$ diffusion space, evaluated using a confidence-scoring network. \citep{nakata2023end} approached the diffusion problem in $R^{3}$, incorporating equivariant constraints across the entire space and employing an equivariant graph neural network for scoring. \citep{qiao2022dynamic} introduced contact prediction modules to direct the molecular diffusion process through the $R^{3}$ manifold, assessing sample plausibility with invariant point attention.

\textbf{Informed Diffusion for Directed Docking}. Most existing diffusion-based methods, including the ones discussed above, do not explicitly consider the precise location of the binding cavity. Despite the technical advancements in blind diffusion docking, these methods have not yet achieved the same level of effectiveness as traditional approaches. This gap has opened opportunities for developing methods that incorporate elements of traditional docking. To the best of our knowledge, the only method integrating the $\mathbb{T}^3 \times SO(3) \times SO(2)^m$ diffusion process with targeted guidance to specific protein surface points is DiffDock-Pocket\footnote{Referred to as DD-Pocket for the remainder of the manuscript.} \citep{plainer2023diffdock}. This method not only adopts a more traditional approach to docking but also extends it by allowing flexibility in side chains.

\section{Methods}

Our approach builds on the concept prevalent in traditional docking strategies, where specific geometric shapes like spheres or boxes demarcate potential binding areas, excluding the remainder of the protein structure. We refine this concept in DiffDock by directing incremental updates during the denoising procedure toward a targeted binding region and conformation.

We utilize a guided diffusion strategy, drawing inspiration from \citet{guideddiff}, by introducing a guiding vector $V_{guide}$ in place of a conventional trained classifier. This vector integrates domain expertise into the diffusion process by altering the update mechanism. The intensity of these alterations is controlled by the hyperparameter $\gamma$, allowing for dynamic adjustments based on the proximity and direct route to the target regions. This method ensures that the direction and magnitude of $V_{guide}$ are precisely aligned with the distance and shortest path to the designated areas, optimizing the diffusion trajectory towards the desired binding site.

\begin{equation} \label{pseudoupdate}
    V_{update} = (1-\gamma) V_{DiffDock} + \gamma V_{guide}
\end{equation}
where
\begin{equation} \label{guidingvector}
    V_{guide} = \alpha \cdot v_{dir}
\end{equation}
being $v_{dir}$ the vector tangent to the shortest path and $\alpha$ the distance towards the selected center or boundary region.  For a detailed analysis of the range of possible $\gamma$ values, we refer the reader to Appendix \ref{gamma_ben}.

Although implementing this method in $R^3$ would be straightforward, DiffDock operates in a more complex space, $P = \mathbb{T}^3 \times SO(3) \times SO(2)^m$, to accurately reflect the degrees of freedom involved in molecular docking. Consequently, we calculate the shortest paths and distances within each component of this product space, effectively tailoring the diffusion process to this task.

\subsection{Boundaries and Geodesics}

Shortest paths in $\mathbb{T}^3$ can easily be determined by leveraging straight lines and the geometry of a hypothetical sphere to integrate binding pocket information effectively. This approach helps mitigate the risk of selecting incorrect docking pockets, a prevalent issue in blind docking strategies that often leads to out-of-distribution errors \citep{blind-gz}. The incorporation of this spherical model is instrumental in refining the selection process, enhancing the algorithm's ability to direct the denoising process to given docking sites.

However, the challenge of ensuring the physical realism of docking poses—highlighted by concerns over steric clashes and distorted bond angles \citep{posebysters}—necessitates further refinement of our model. To address these issues, our method extends to the $SO(3)$ and $SO(2)^m$ spaces, where we employ geodesics to define the shortest paths. This approach is critical for accurately guiding our diffusion updates with $V_{guide}$ vectors, ensuring that the docking poses are not only theoretically viable but also physically plausible. For an in-depth explanation of how $V_{guide}$ vectors are defined and integrated into the diffusion process, please refer to Appendix \ref{alg:geod}.

\subsection{Soft Constraints} \label{meth:soft}
To balance the precision and flexibility of DiffDock, we implement measures to modulate the influence of guidance vectors, recognizing their potential to alter the model's learned dynamics and impact performance within designated spatial regions. We introduce a dynamic adjustment of the guidance vector's influence through a decreasing sigmoid schedule for the $\gamma$ parameter, progressively diminishing its impact through the diffusion process. This ensures the initial guidance by expert knowledge gradually cedes to the model's intrinsic optimization capabilities.

Moreover, we enable the consideration of multiple potential regions within the $SO(3)$ and $SO(2)^m$ spaces, guiding updates towards the nearest boundary. Once within a targeted region, DiffDock is allowed autonomy in refining rotation and torsion angles, unencumbered by external guidance. This dual-phase optimization strategy—global optimization guided by expert knowledge in early steps, followed by local, autonomous refinement by DiffDock—ensures a balance between adherence to biologically plausible paths and the discovery of optimal docking configurations.

\subsection{Evaluation}
To facilitate high-throughput evaluation without direct expert input for each sample, we employ a fuzzing strategy, adjusting true labels to simulate expert knowledge. This involves creating guidance spheres for translation with a 7Å radius around the ligand's actual center, directing vectors toward these spheres' peripheries, and ceasing guidance upon entrance. For rotation and torsion, we define regions around true angles, adjusted by a fuzzing factor $\eta$ of 0.15, guiding vectors towards these adjusted regions and turning off the guidance when they enter. This fuzzing approach, robust across radius and $\eta$ settings, allows for an effective approximation of expert-directed docking, as detailed in Appendix \ref{fuzz}. Moreover, we perform an ablation test of different initial gamma values in Appendix \ref{gamma_ben} and test the generalization capabilities of our algorithm in Appendix \ref{Dockgen_benchmark}.

\section{Experiments}

To assess our approach, we employ the testing set from PDBBind proposed in \citet{stark2022equibind}, applying a three-fold evaluation strategy. Firstly, we analyze docking pose RMSD, focusing on the Top1 and Top5 poses by confidence for both Apo and Holo structures, and introduce Mean Square Error (MSE) as an additional metric to evaluate errors in rotation states and torsion angles, detailed in Appendix \ref{exp:mse}. Secondly, we examine the physical plausibility of these docking poses using the Posebusters suite \citep{posebysters}, providing insights into the realism of the predicted conformations. Lastly, we conduct an angle transfer test employing maximal common substructure docking to assess our method's ability to generalize across different molecular configurations.

\subsection{Evaluation of docking poses} \label{exp:rmsd}

We evaluated GeoDirDock (GDD) against DiffDock across various setups by comparing the RMSD of generated poses to crystal structures and examining the convergence based on the number of denoising steps taken.

Initially, we tested GDD with translation guidance only (GDD-TR), mimicking conventional docking's focus on specific geometric volumes. Results showed GDD-TR surpassing DiffDock across all metrics, achieving lower RMSD values in fewer steps and maintaining performance across both Apo and Holo receptor configurations, suggesting that targeted guidance enhances docking accuracy and efficiency (Table \ref{tab:docking_methods}).

\begin{table}[h]
\centering
\caption{RMSD performance comparison for Holo and Apo settings across docking methods, showing Top-1 and Top-5 prediction accuracies. Best performances are highlighted in \textbf{bold}. Best performances with only translation guidance are stated in \textit{italic}. The number of denoising steps and the number of samples generated are stated as (steps-samples). Results marked with an asterisk (*) were obtained from \citet{plainer2023diffdock}}
\begin{tabularx}{\textwidth}{l|*{8}{>{\centering\arraybackslash}X}}
\hline
                & \multicolumn{4}{c|}{Holo}                              & \multicolumn{4}{c}{Apo}                               \\ 
                & \multicolumn{2}{c|}{Top-1 RMSD}      & \multicolumn{2}{c|}{Top-5 RMSD}      & \multicolumn{2}{c|}{Top-1 RMSD}      & \multicolumn{2}{c}{Top-5 RMSD}      \\ 
                & \%$<$2       & Med      & \%$<$2       & Med      & \%$<$2      & Med     & \%$<$2      & Med     \\ \hline
SMINA* (rigid)        & 32.5 & 4.5 & 46.4 & 2.2 & 6.6  & 7.7 & 15.7 & 5.6 \\
GNINA* (rigid)         & 42.7 & 2.5 & 55.3 & 1.8 & 9.7  & 7.5 & 19.1 & 5.2 \\\hline
DiffDock (10-10) & 34.19       & 3.53          & 40.17       & 2.44          & 27.14      & 4.62         & 36.28      & 3.09         \\
DiffDock (20-10) & 37.08       & 3.50          & 44.66       & 2.60          & 27.51      & 4.56         & 38.40      & 2.85         \\
DiffDock (20-40) & 38.27       & 3.12          & 46.65       & 2.15          & 27.01      & 4.85         & 37.93      & 3.22         \\ 
DD-Pocket*  (20-10)    & 47.7 & 2.1 & 56.3 & 1.8 & 41.0 & 2.6 & 47.6 & 2.2 \\
DD-Pocket*  (20-40)    & 49.8 & 2.0 & 59.3 & 1.7 & 41.7 & 2.6 & 47.8 & 2.1 \\\hline
GDD-TR (10-10)   & 41.62       & 2.69          & 48.88       & 2.03          & 31.71      & 3.62         & 42.00      & 2.50         \\
GDD-TR (20-10)   & 44.97       & 2.37          & 50.56       & 1.94          & 34.29      &  \textit{3.21}       & 44.00      & 2.27         \\
GDD-TR (20-40)   & \textit{48.88}     &  \textit{2.05}          &  \textit{57.82} &  \textit{1.70} &  \textit{34.38}      & 3.33         &  \textit{47.28}      &  \textit{2.11}         \\ \hline
GDD-Full (10-10) & 63.97       & 1.52          & 67.60       & 1.31          & 51.29      & 1.95         & 58.74      & 1.60\\
GDD-Full (20-10) & 68.44       & 1.24          & 70.11       & 1.16          & \textbf{59.43}     & 1.60         & \textbf{65.14}      & 1.43\\
GDD-Full (20-40) & \textbf{68.72} & \textbf{1.22} & \textbf{71.51}       & \textbf{1.14}          & 58.86 & \textbf{1.59} & 63.71 & \textbf{1.38} \\ \hline
\end{tabularx}
\label{tab:docking_methods}
\end{table}

Upon extending guidance to include torsions and rotations, imposing a more substantial prior than translation alone, we anticipated improvements in the reconstruction of crystallographic structures. GDD-Full, informed across all three spaces, markedly outperformed both GDD-TR and DiffDock, recovering approximately 68\% of poses within 2Å of the reference, significantly higher than with translation-only guidance. This comprehensive approach not only improved pose accuracy but also maintained the trend of achieving optimal results in fewer steps, highlighting the efficacy of our method in enhancing docking precision and computational efficiency. Detailed MSE analysis for torsion and rotation are available in Appendix \ref{exp:mse}.

\subsection{Physical plausibility of informed diffusion poses}\label{exp:posebusters}

\begin{table}[h]
\centering
\caption{Assessment of Posebusters scores for Holo and Apo scenarios, comparing docking structure and re-docking accuracies across DiffDock and GDD methods. Best performances are highlighted in \textbf{bold}. The number of denoising steps and the number of samples generated are stated as (steps-samples).  Results marked with an asterisk (*) were obtained from \citet{plainer2023diffdock}}
\begin{tabularx}{\textwidth}{l|*{8}{>{\centering\arraybackslash}X}}
\hline
                & \multicolumn{2}{c}{Holo}                              & \multicolumn{2}{c}{Apo}                               \\ 
                & \multicolumn{1}{c}{Docking Structure}      & \multicolumn{1}{c|}{Re-docking}      & \multicolumn{1}{c}{Docking Structure}      & \multicolumn{1}{c}{Re-docking}      \\  \hline
DiffDock (10-10) &  23.13   &   15.31 &   7.06 & 4.12                   \\
DiffDock (20-10) &  25.84  &  16.11  &  7.43  &   4.0                 \\
DiffDock (20-40) &   26.67  &  16.0 &   14.45 & 4.05                    \\ 
DD-Pocket (20-40)* &  29.4 &  17.4  &    \textbf{21.6}  &  \textbf{10.9}   \\ \hline
GDD-TR (10-10)   &   23.00  &  15.33 &  3.43 &  1.71                   \\
GDD-TR (20-10)   &   20.00 &  15.33 &   10.29 & 6.29                 \\
GDD-TR (20-40)   &   22.00  & 14.33 &   9.14 &  5.14        \\ \hline
GDD-Full (10-10) &   29.33  & 24.33 &  8.62  & 6.90                \\
GDD-Full (20-10) &  28.00  & 24.00  &  12.0   &  10.86            \\
GDD-Full (20-40) &  \textbf{30.67} &  \textbf{26.67}  &   11.43  &  9.71          \\ \hline

\end{tabularx}
\label{tab:posebusters}
\end{table}

The Posebusters evaluation (Table \ref{tab:posebusters}) reveals that GDD-Full outperforms GDD-TR and DiffDock in Holo structures, demonstrating its superior accuracy in docking and re-docking tasks. This success highlights GDD-Full's ability to align with the physical aspects of molecular docking, attributed to its comprehensive guidance across multiple spaces.

In contrast, GDD-TR suffers a performance degradation in both Holo and Apo structures.
This aligns with our hypothesis of full guidance being needed to reproduce physically plausible poses, as the main objective of translational guidance is the correct selection of the binding pocket, not the increment of physical plausibility of poses.

\subsection{Angle transfer as maximal common substructure docking}\label{exp:angletransf}
In the lead optimization phase of drug discovery, template-based modeling is crucial for assessing binding affinity changes among chemically similar compounds \citep{template_modelling}. This approach involves using a single compound as a reference template to maintain consistent a shared topology while optimizing the distinct elements of each molecule.

To evaluate our method's capabilities beyond self-docking and informed directed docking, we decided to benchmark GDD as a potential tool for template-based modeling. For this we selected crystallized BACE structures from the D3R Grand Challenge 4 \citep{D3R4}, employing previously identified templates for angle transfer. 

We conducted an MCS search to transfer torsion angles associated with heavy atoms common between the template and target molecules, leaving other angles uninformed. This focused evaluation, termed GDD-Tor, exclusively examines the impact of torsion angle transfer, deliberately omitting guidance in other spatial dimensions to isolate the effects of this specific mechanism.

\begin{table}[h]
\centering
\caption{Comparison of MCS Docking performance, showing RMSD and MSE metrics for Top-1 and Top-5 predictions between DiffDock and GDD-Tor methods. The number of denoising steps and the number of samples generated are stated as (steps-samples).}
\begin{tabularx}{\textwidth}{l|*{8}{>{\centering\arraybackslash}X}}
\hline
                & \multicolumn{2}{c|}{Top-1 RMSD}      & \multicolumn{2}{c|}{Top-5 RMSD}      & \multicolumn{2}{c|}{Top-1 MSE}      & \multicolumn{2}{c}{Top-5 MSE}      \\ 
                & \%$<$2       & Med      & \%$<$2       & Med      & Avg    & Med     & Avg     & Med     \\ \hline
DiffDock (20-40) &    50.0    &   1.93       &  55.0   &   1.71      &  2.33   &  2.30      &  1.37    &  1.45       \\ \hline

GDD-Tor (20-40)   &  63.2  &  1.76   & 63.2 & 1.69 &  1.79   &  1.35    & 1.31   & 1.21        \\ \hline
\end{tabularx}
\label{tab:mcs} 
\end{table}

As we can determine by the results in Table \ref{tab:mcs}, this angle transfer already marks an improvement both in RMSD and torsion MSE. These preliminary results outline great potential for our method and we look to further develop this application and corresponding benchmarks.

\section{Conclusions and Future Work}
In this study, we introduce GeoDirDock, a novel framework that enhances generative docking models through geodesic guidance. GeoDirDock significantly improves docking precision and boosts the physical plausibility of results compared to traditional blind docking methods like DiffDock. By incorporating guidance across translation, torsion, and rotation dimensions, GeoDirDock outperforms both DiffDock and its translation-only variant, achieving precise docking poses with fewer computational steps. These results pose guided docking as a promising avenue for AI-enabled molecular docking, ensuring both efficiency and accuracy in molecular docking.

Future work would focus on improving in generalizability to completely unseen ligand and protein chemistries, thus, postulating GeoDirDock as a valuable tool for prospective docking campaigns. We are also keen on extending this algorithm towards including protein flexibility in the docking procedure, aiming for a more realistic setting of the docking protocol. The inclusion of more realistic priors, in both backbone and side chain angles, would solve possible steric clashes between ligands and proteins, performing implicit induced fit docking.

\bibliography{iclr2024_conference}
\bibliographystyle{iclr2024_conference}
\newpage
\appendix

\section{Algorithms and computations} \label{alg:geod}

\subsection{Geodesics in $\mathbb{T}^3$}
Geodesics in $\mathbb{T}^3$ can be represented as straight lines in $R^3$. 
We parameterize a binding sphere B with a center $\epsilon$ and a radius $r$ and $p_t$ is defined as the ligand's center of mass. Then, the direction vector of translation guidance can be computed as:

\begin{equation}
v_{dir}^{tr}=
\begin{cases}
    0 & \text{if } p_{t} \in  B(\epsilon;r)\\
    \frac{\epsilon - p_{t}}{\|\epsilon - p_{t}\|_2}& \text{if not}
\end{cases}
\end{equation}
We choose the scaling factor $\alpha^{tr}$, due to empirical performance, to be:
\begin{equation}
    \alpha^{tr} = \|\epsilon - p_{t}\|_2
\end{equation}

\begin{figure}[htbp]
    \centering
    \includegraphics[scale=0.355]{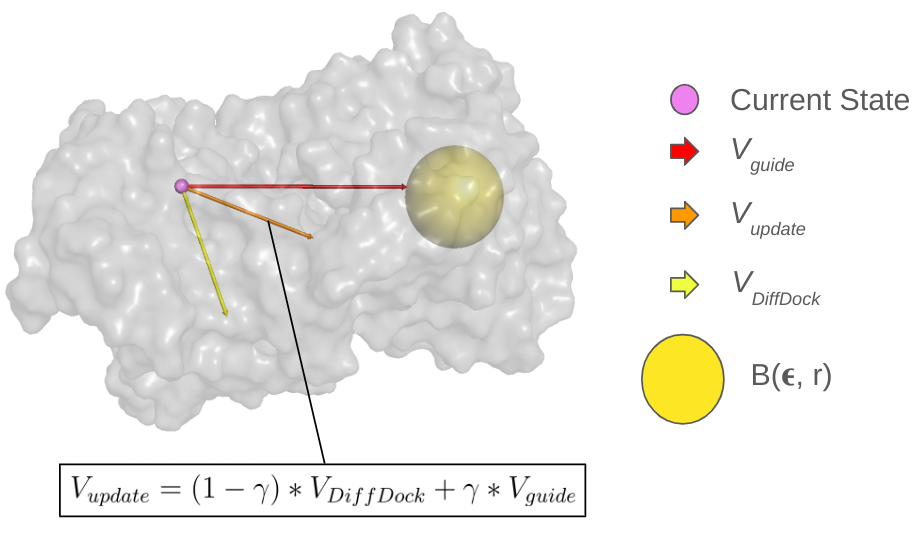}
    \caption{Guidance in $\mathbb{T}^3$ for a sample sphere in \textit{1a0q}.}
    \label{fig:molT3}
\end{figure}

\subsection{Geodesics in $SO(3)$}

\begin{figure}[htbp]
    \centering
    \includegraphics[scale=0.355]{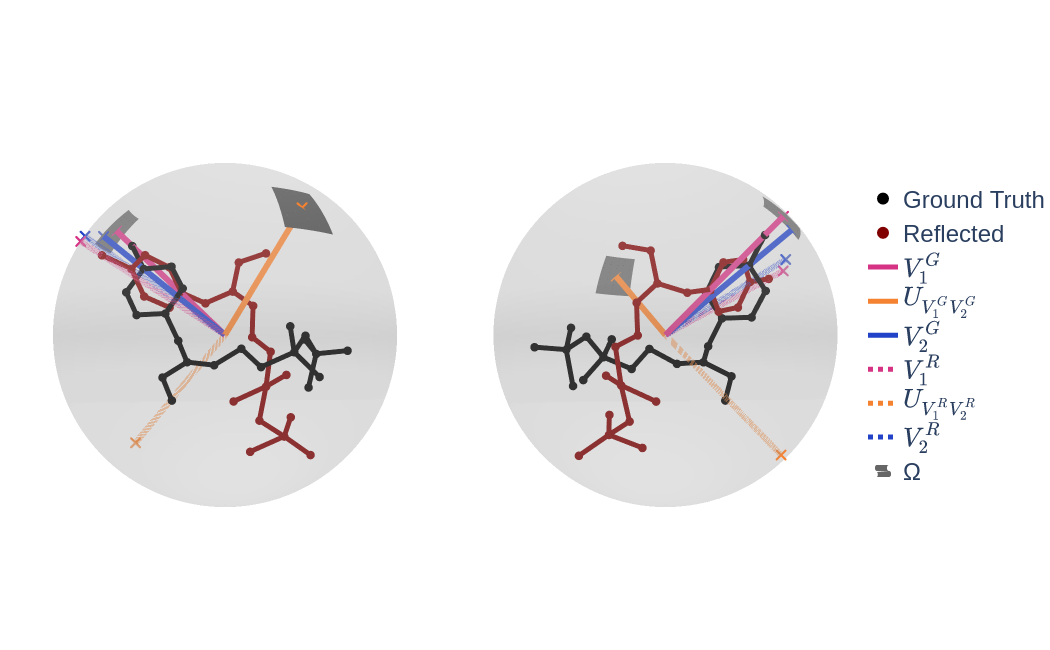}
    \caption{Rotation state vectors in $S^2$ for a reflected molecule versus its ground truth.}
    \label{fig:molSO3}
\end{figure}

Rigid rotations of the ligand are defined as rotations around its center of mass , corresponding to the its orientation within the pocket \citep{Corso2022DiffDockDS}. As this is extremely complicated to define a-priori by an expert, we choose to cast this problem onto finding the optimal rotation states of said center with respect to its atoms. For this, we utilize the axis formed between the center of masses and the first atom as our prior, and to account for possible reflections of the plane formed by this vector and the DiffDock update, we will also include the axis formed by the orthogonal vector of said axis and the one formed between the center of mass and the second atom \ref{fig:molSO3}. Those two vectors will be the axes of rotation for which we aim to optimize the rotation states.

A unit 2-sphere embedded in $R^3$ can be denoted as $S^2 \subset R^3$ and in this sphere, we can encode the rotation states of our target using polar coordinates:

\begin{equation}
r_{t} = \phi \in [0,2\pi), \theta \in [0,\pi)
\end{equation}

We can therefore define the selected regions of interest for the rotation, $\Omega$, as:

\begin{equation}
\Omega = (\bigsqcup_{i=1}^p [\phi_{1}^{i},\phi_{2}^{i}]) \times (\bigsqcup_{i=1}^{th} [\theta_{1}^{i},\theta_{2}^{i}])
\end{equation}

 For the sake of simplicity, although our implementation supports this more general definition with multiple regions, we will define $\Omega$ as a unique region for each of the axes:
 
 \begin{equation}
 \Omega = [\phi_{1},\phi_{2}]\times[\theta_{1},\theta_{2}]
 \end{equation}
 
We parameterize both rotations as points on this unit sphere and because we need to compute the guiding vector with the condition of being tangent to the geodesic, we map our points as vectors $v \in R^3 $, according to their position in the sphere, as:

\begin{equation}
    v = (\cos{\phi}\sin{\theta}, \sin{\phi}\sin{\theta}, \cos{\theta})
\end{equation}

For the corresponding $\phi$ and $\theta$ of each point, obtaining $v_\alpha$ and $v_\beta$.  

We are interested in both of the geodesics that cover the shortest distance between each pair of two points. Geodesics on the sphere can be defined as the great circle centered in the origin, with the plane containing it having the orthogonal basis $v_\alpha$ and $u$, defined as:

\begin{equation}
    g(t) = \cos(t) v_{\alpha} + \sin(t) u 
\end{equation}
With $u$ defined as:
\begin{equation}
    u = \frac{v_{\beta} - (v_{\beta}^{t} \cdot v_{\alpha})v_{\alpha}}{\|v_{\beta} - (v_{\beta}^{t} \cdot v_{\alpha})v_{\alpha}\|}
\end{equation}

Point $\beta$ can then be reached with:

\begin{equation}
    t^{*} = \arccos(v_{\beta}^{t} \cdot v_{\alpha})
\end{equation}

With one $t^{*}$ defined for each selected axis, as $t^{*}_1$ and $t^{*}_2$.

We cannot directly add these vectors to the DiffDock update, $v_{r_{t}}$, so we choose to sequentially apply the rotation updates one by one. 

\begin{equation}
v_{dir}^{rot}=
\begin{cases}
    0 & \text{if } r_{t} \in  [\phi_{1},\phi_{2}]\times[\theta_{1},\theta_{2}]\\
    t^{*}_2 \cdot ((t^{*}_1 \cdot (v_{r_{t}} \times u_1)) \times u_2) & \text{if not}
\end{cases}
\end{equation}

The scaling factor was set as 1, as we found poor empirical performance when varying it.

\begin{equation}
    \alpha^{rot} = 1
\end{equation}

\begin{figure}[htbp]
    \centering
    \includegraphics[scale=0.155]{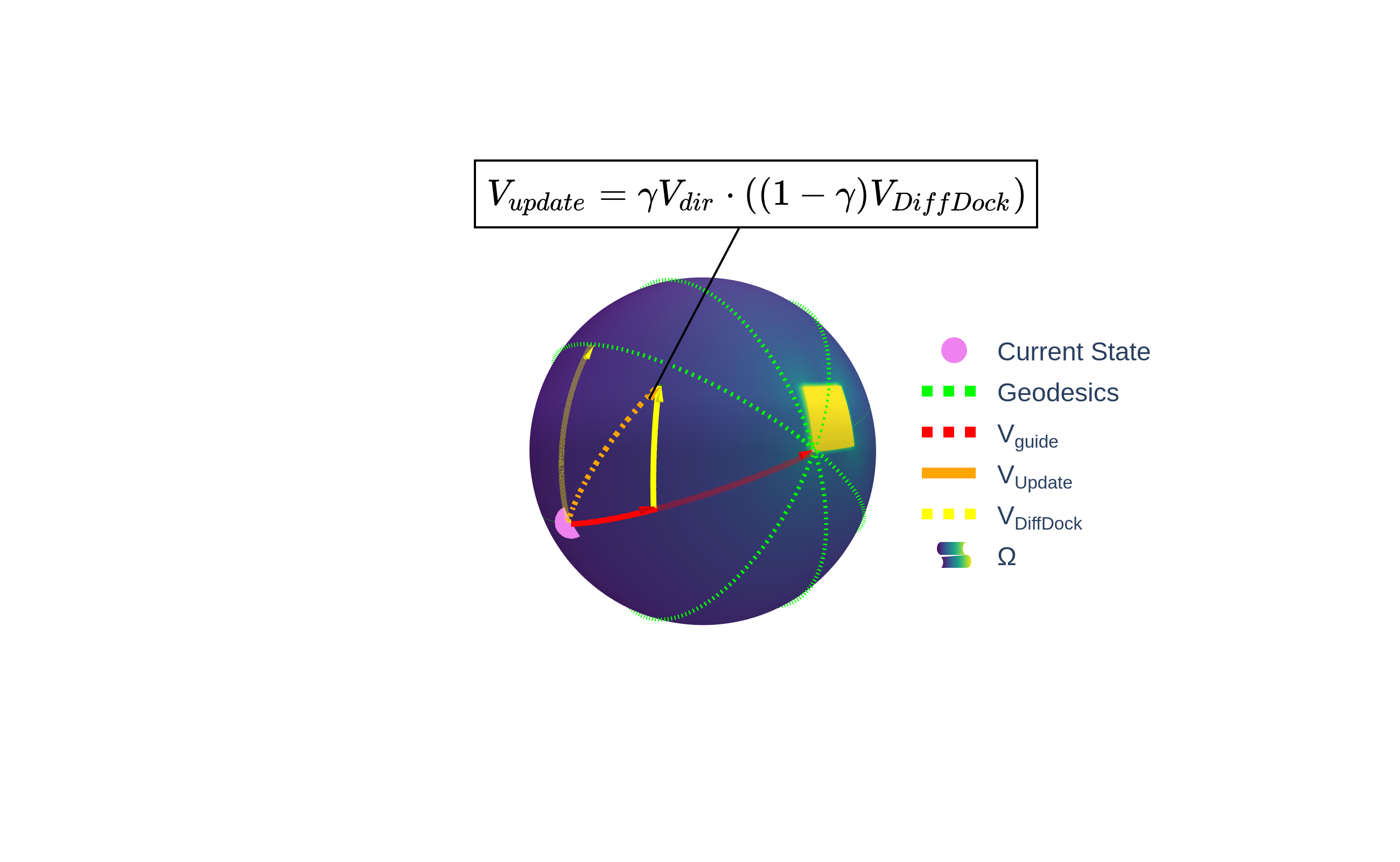}
    \caption{$SO(3)$ guidance in $S^2$ for a sample rotation state.}
    \label{fig:gSO3}
\end{figure}

\subsection{Geodesics in $SO(2)^m$}

\begin{figure}[htbp]
    \centering
    \includegraphics[scale=0.175]{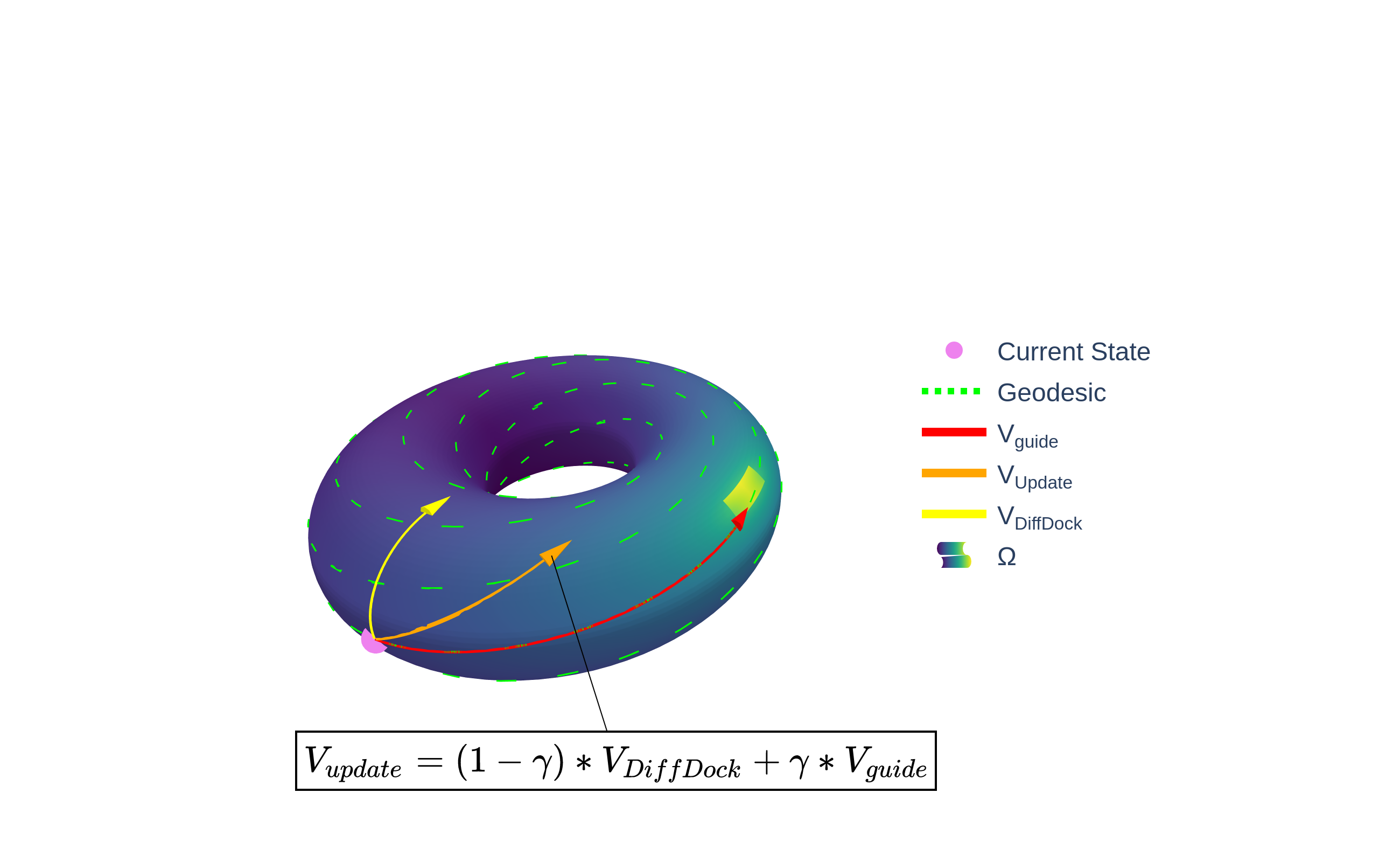}
    \caption{Guidance in $SO(2)^m$ for a 2 torsion angle compound.}
    \label{fig:torus}
\end{figure}

As mentioned in \citet{Corso2022DiffDockDS}, $SO(2)^m$ is diffeomorphic to the hypertorus $\mathcal{T}^m$, therefore the guidance calculations will be performed in this m-dimensional hypertorus. Recalling the definition of the torus being the product space of two 1-spheres or circles:
\begin{equation}
    \mathcal{T}^2 = S^1 \times S^1
\end{equation}
and one-dimensional torus being equal to a circle $\mathcal{T}^1 = S^1$, we can define the hypertorus $\mathcal{T}^m$ as the product space of m 1-spheres:

\begin{equation}
    \mathcal{T}^{m} = \underbrace{\mathcal{S}^{1} \times \ldots \times \mathcal{S}^{1}}_{m}
\end{equation}

With $\mathcal{T}^m$ being defined as the following quotient space:
\begin{equation}\label{quotientspace}
    \mathcal{T}^{m} \cong  \mathcal{R}^{m} /\mathcal{Z}^{m} \cong \mathcal{R}^{m} / 2\pi \mathcal{Z}^{m}
\end{equation}

With this information, we can determine geodesics in the hypertorus to be straight lines in the hypercube that reappear in the diametrically opposite point when they collide with a boundary. Leveraging this knowledge, we now can compute the desired shortest path geodesic, its length and the tangent vector.

Given two points $z_1, z_2 \in \mathcal{T}^{m}$, let $\mathcal{G}(z_{1},z_{2})$ be the set of all possible geodesics that join $z_{1}$ and $z_{2}$. We are interested in finding the shortest-path geodesic between $z_{1}$ and $z_{2}$:
\begin{equation}
     \mathbb{g}_{z_{1},z_{2}} = argmin_{g \in \mathcal{G}(z_{1},z_{2})} l(g(z_{1},z_{2}))
\end{equation}

With its length being defined as $l(\mathbb{g}_{z_{1},z_{2}})$, and the tangent vector being defined as $v_{dir}(\mathbb{g}_{z_{1},z_{2}})$.
We will define the selected regions, $\Omega$ as:
\begin{equation}
 \Omega = (\bigsqcup_{k=1}^{r_{1}}[\theta_{2k-1}^{1},\theta_{2k}^{1}]) \times \ldots \times (\bigsqcup_{k=1}^{r_{m}}[\theta_{2k-1}^{m},\theta_{2k}^{m}]).
\end{equation}
Being each set of $\theta_{2k-1},\theta_{2k}$ the delimiters of the region in each of the 1-spheres composing the torus.

Although we build our algorithms to be able to handle multiple regions per angle, we will omit this from the algorithms for simplicity. $\Omega$ is therefore defined as the unique region of interest in the hypertorus. As expressed in \ref{meth:soft}, we need to determine if our current state is within the bounds of the selected region to turn off the guidance: 
\begin{algorithm}[H]
\caption{Is $z \in \mathcal{T}^{k}$ in $Proj_{\mathcal{T}^{k}}\Omega$?}\label{inO}
\hspace*{\algorithmicindent} \textbf{Input:} $0 < n \leq m, \{j_{1},\ldots j_{n}\} \subset \{1,\ldots,m\},z \in \mathcal{T}^{n},\Omega$ \\
    \hspace*{\algorithmicindent} \textbf{Output: True or False} 
\begin{algorithmic}
\State $count \gets 1$
\For{$i$ in $\{j_{1},\ldots,j_{n}\}$}
    \If{$\theta_{2}^{i} \geq \theta_{1}^{i}$}
        \If{not $\theta_{1}^{i} \leq z_{count}$ or not $z_{count} \leq \theta_{2}^{i}$} 
            \State \Return False
        \EndIf
    \Else
        \If{not $z_{count} \geq \theta_{1}^{i}$ and not $z_{count} \leq \theta_{2}^{i}$} 
            \State \Return False
        \EndIf
    \EndIf
    \State $count \gets count + 1$
\EndFor
\State \Return True
\end{algorithmic}
\end{algorithm}
We will also need to compute the vector tangent to the geodesic between the points and the length of said geodesic:
\begin{algorithm}[H]
\caption{Computation of tangent vector $v_{dir}(\mathbb{g}_{z_1,z_2})$ and l($\mathbb{g}_{z_1,z_2}$)}\label{alg1}
\hspace*{\algorithmicindent} \textbf{Input:} $m > 0, z_{1}, z_{2} \in \mathcal{T}^{m}$ \\
    \hspace*{\algorithmicindent} \textbf{Output: $l(\mathbb{g}_{z_1,z_2}) \in \mathcal{R}_{\geq 0}$, $v_{dir}(\mathbb{g}_{z_1,z_2}) \in \partial B(0;1) \subset \mathcal{R}^{m}$} 
\begin{algorithmic}
\State $l \gets 0$
\For{$i$ in $\{1,\ldots,m\}$}
    \If{$z_{2}^{i} \geq z_{1}^{i}$}
        \If {$z_{2}^{i} - z_{1}^{i} \leq z_{1}^{i} + (2\pi - z_{2}^{i})$}
            \State $l \gets l + (z_{2}^{i} - z_{1}^{i})^{2}$ 
            \State $v_{dir}^{i} \gets z_{2}^{i} - z_{1}^{i}$
        \Else
            \State $l \gets l + (z_{1}^{i} + (2\pi - z_{2}^{i}))^{2}$
            \State $v_{dir}^{i} \gets z_{2}^{i} - (2\pi + z_{1}^{i})$
        \EndIf
    \Else 
        \If{$z_{1}^{i} - z_{2}^{i} \leq (2\pi - z_{1}^{i}) + z_{2}^{i}$}
            \State $l \gets l + (z_{1}^{i} - z_{2}^{i})^{2}$
            \State $v_{dir}^{i} = z_{2}^{i} - z_{1}^{i}$
        \Else 
            \State $l \gets l + ((2\pi - z_{1}^{i}) + z_{2}^{i})^{2}$
            \State $v_{dir}^{i} = (z_{2}^{i} + 2\pi) - z_{1}^{i}$
        \EndIf
    \EndIf
\EndFor
\State $l \gets \sqrt{l}$
\State $v_{dir} \gets \frac{1}{l}(v_{dir}^{1},\ldots,v_{dir}^{m})$
\end{algorithmic}
\end{algorithm}
In this computation, we have to take into account which of the boundaries is closer to our current state, and which direction is closer in the torus geometry. Therefore our final boundary vector $\tau$ can be computed as:

\begin{algorithm}[H]
\caption{$argmin_{\tau \in \partial \Omega}\min{l(\mathbb{g}_{x_{t},\tau_{t}})}$}\label{minTau}
\hspace*{\algorithmicindent} \textbf{Input:} $m \geq 0, x_{t}, \Omega$ \\
    \hspace*{\algorithmicindent} \textbf{Output: $\tau$} 
\begin{algorithmic} 
\For{$i$ in $\{1,\ldots,m\}$}
    \If{$x_{t}^{i} \in [\theta_{1}^{i},\theta_{2}^{i}]$} \Comment{$x_{t}^{i} \in [\theta_{1}^{i},\theta_{2}^{i}]$ = Alg\ref{inO}($n= 1,m,\{i\},x_{t}^{i},\Omega$)}
        \State $\tau^{i} \gets x_{t}^{i}$
    \Else
        \If{$\min{l(\mathbb{g}_{x_{t}^{i},\theta_{1}^{i}})}\leq \min{l(\mathbb{g}_{x_{t}^{i},\theta_{2}^{i}})}$}
            \State $\tau^{i} \gets \theta_{1}^{i}$
        \Else \State  $\tau^{i} \gets \theta_{2}^{i}$ 
        \EndIf \Comment{$l(\mathbb{g}_{x_{t}^{i},\theta_{k}^{i}})$ from Alg\ref{alg1}($1,x_{t}^{i}, \theta_{k}^{i}$)}
    \EndIf
\EndFor
\State $\tau \gets (\tau^{1},\ldots,\tau^{m})$
\end{algorithmic}
\end{algorithm}
This vector indicates the closest boundaries of our selected region to our current state, and our subsequent guiding vector can be computed using Algorithm \ref{alg1}.
\begin{equation}
    v_{dir}^{tor} = 
\begin{cases}
    0 & \text{if } x_{t} \in \Omega \\
    v_{dir}(\mathbb{g}_{x_t,\tau}) & \text{if not}
\end{cases}
\end{equation}

We choose the $\alpha$ scaling factor to be the length of the shortest path geodesic, also computed with Algorithm \ref{alg1}:
\begin{equation}
    \alpha^{tor} = l(\mathbb{g}_{x_t,\tau})
\end{equation}

\section{Evaluation via rotational and torsional MSE}\label{exp:mse}
In this section, we introduce the usage of Mean Square Error in torsion and rotation as a measure that decouples distance to the binding site and bond and torsion errors. 

We believe this separation to be beneficial in a high granularity analysis, as errors in binding pocket detection can easily affect the ligand RMSD measurements while hiding a correct assignment of torsion angles. So, we further analyze the previously obtained poses from Section \ref{exp:rmsd} in rotation and torsion MSE.
\begin{table}[h]\label{res:mse}
\centering
\caption{Comparison of MSE in rotation and torsion across docking methods.}
\begin{tabularx}{\textwidth}{l|*{8}{>{\centering\arraybackslash}X}}
\hline
                & \multicolumn{4}{c|}{Holo}                              & \multicolumn{4}{c}{Apo}                               \\ 
                & \multicolumn{2}{c|}{Top-1 MSE}      & \multicolumn{2}{c|}{Top-5 MSE}      & \multicolumn{2}{c|}{Top-1 MSE}      & \multicolumn{2}{c}{Top-5 MSE}      \\ 
                 & Rot     & Tor      & Rot     & Tor     & Rot       & Tor    & Rot     & Tor    \\ \hline
DiffDock (10-10) & 1.82    & 2.35     & 0.32    & 1.43    & 1.77      & 2.37   & 0.40    & 1.30      \\
DiffDock (20-10) & 1.66    & 2.22     & 0.26    & \textit{1.25}    & 2.13      & 2.38   & 0.47    & 1.33     \\
DiffDock (20-40) & 1.58    & 2.17     & \textit{0.20}    & 1.40    & \textit{1.32}      & 2.31   & 0.40    & 1.40        \\ \hline
GDD-TR (10-10)   & 1.64    & 2.28     & 0.27    & 1.30    & 1.82      & 2.48   & 0.35    & 1.37      \\
GDD-TR (20-10)   & \textit{1.38}    & 2.28     & 0.22    & 1.27    & 1.97      & \textit{2.29}   & \textit{0.24}    & \textit{1.32} \\
GDD-TR (20-40)   & 1.58    & \textit{2.17}     &\textit{0.20}    & 1.40    & 1.33      & 2.39   & 0.26    & 1.40  \\ \hline
GDD-Full (10-10) & 0.04    & 0.08     & 0.02    & 0.05    & 0.05      & 0.08   & 0.02    & 0.05    \\
GDD-Full (20-10) & \textbf{0.02}    & \textbf{0.03}     & \textbf{0.01}    & \textbf{0.02}    & \textbf{0.03}      & \textbf{0.03}   & \textbf{0.01}    &\textbf{0.02} \\
GDD-Full (20-40) & \textbf{0.02}    & \textbf{0.03}     & \textbf{0.01}    & \textbf{0.02}    & \textbf{0.03}      & \textbf{0.03}   & \textbf{0.01}    & \textbf{0.02} \\ \hline
\end{tabularx}

\label{tab:mse}
\end{table}

In these results we can clearly determine a positive effect of the full guide on both Holo and Apo structures, supporting its hypothesized usage for correcting bond and rotation state errors. Furthermore, we also detect mild improvements with only translational guides, which leads us to believe that faster convergence upon correct binding sites allows DiffDock to focus in better bond and rotation states, increasing RMSD performance.

\section{Region Fuzzing Ablation} \label{fuzz}
In this section we seek to evaluate the effect of altering the $\eta$ parameter and radius parameter of our algorithm, weakening the prior that we impose upon the translation, rotation and torsion states. We perform this ablation test in a smaller set of the PDBBind testing dataset, a set of 30 complexes randomly sampled. 

In each of the sections, DiffDock will be compared against GDD-full with variations in the fuzzing hyperparameter corresponding to the respective space where the ablation is performed, maintaining default values for the rest of the spaces. The comparison will be performed as the mean and median step-wise error of the 40 samples and 20 steps protocol for both algorithms. 

\subsection{Translation}
For translation we will consider DiffDock base against GDD full with spheres of radius 5,7,10, and 12 \r{A}, 7 \r{A} being the default of our algorithm.

\begin{figure}[H]
\centering
\begin{subfigure}{.5\textwidth}
  \centering
  \includegraphics[width=\linewidth]{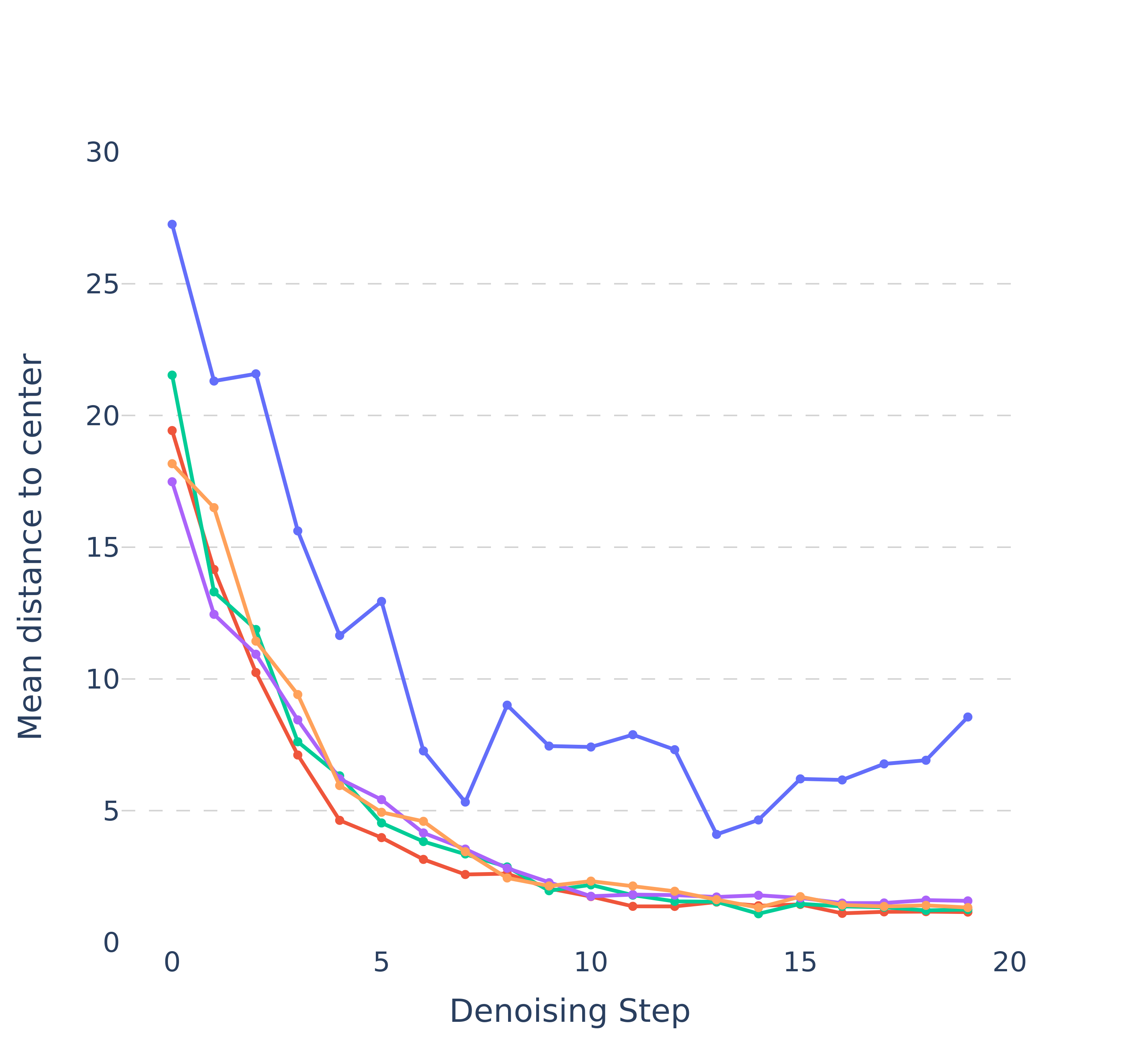}
  \label{fig:trsub1}
\end{subfigure}%
\begin{subfigure}{.5\textwidth}
  \centering
  \includegraphics[width=\linewidth]{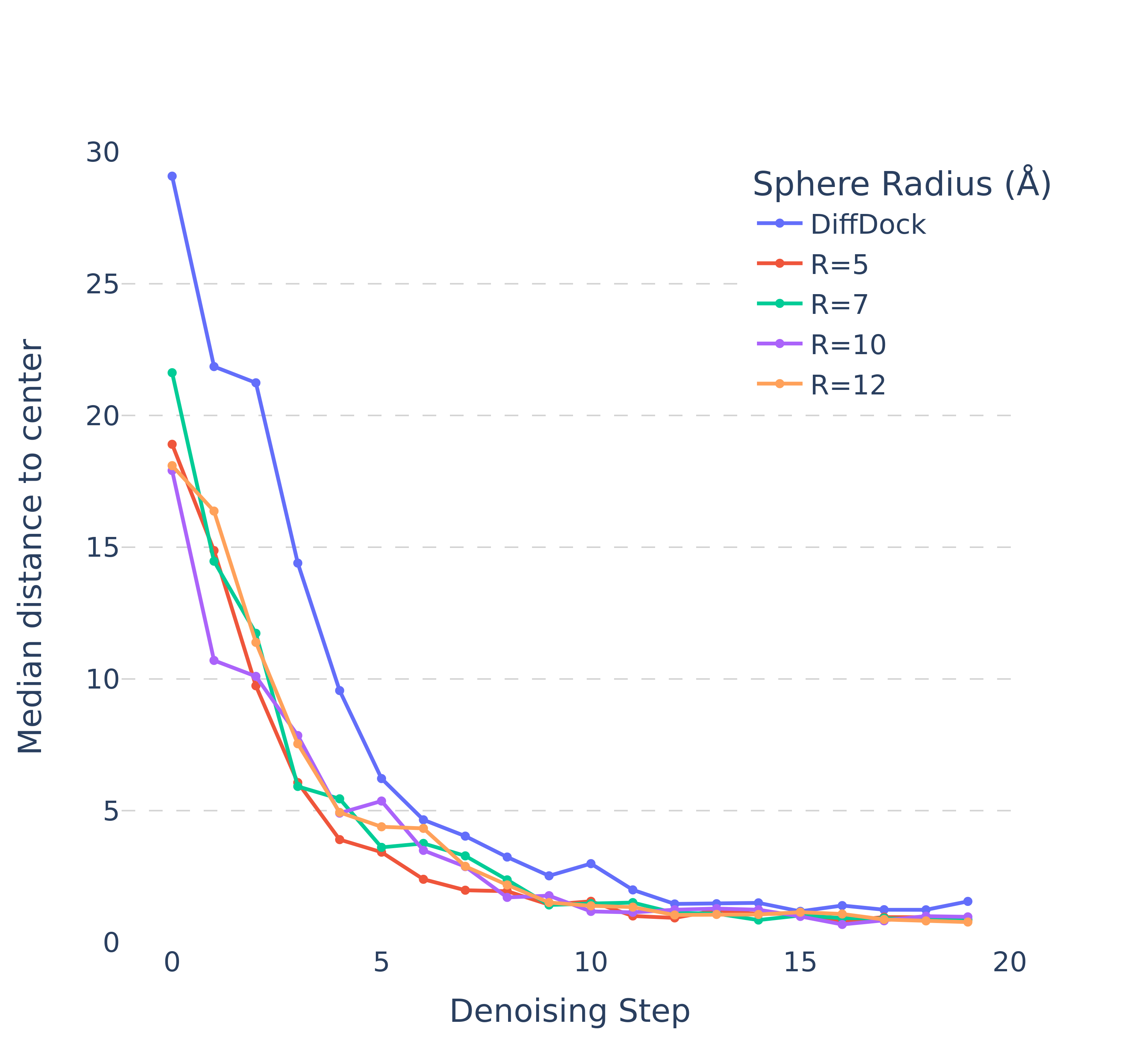}
  \label{fig:trsub2}
\end{subfigure}
\caption{Mean and median distance to center of sphere versus denoising step.}
\label{fig:tr}
\end{figure}

As we are able to determine from Figure \ref{fig:tr}, our algorithm is very robust to the radius parameter, with a rough estimate of a binding pocket being enough for discarding outlier values that heavily affect the mean distance to the correct binding site.

We are also able to observe a faster convergence across the denoising steps, highlighting the potential of our method for reducing the computational cost of DiffDock in high-throughput virtual screening campaigns.

\subsection{Rotation}

In rotation, we set the same default value for both values of $\eta$ for angles $\phi$ and $\theta$ at 0.15. We further evaluated values of 0.3, 0.45 and 0.6 for both $\eta$.

\begin{figure}[H]
\centering
\begin{subfigure}{.5\textwidth}
  \centering
  \includegraphics[width=\linewidth]{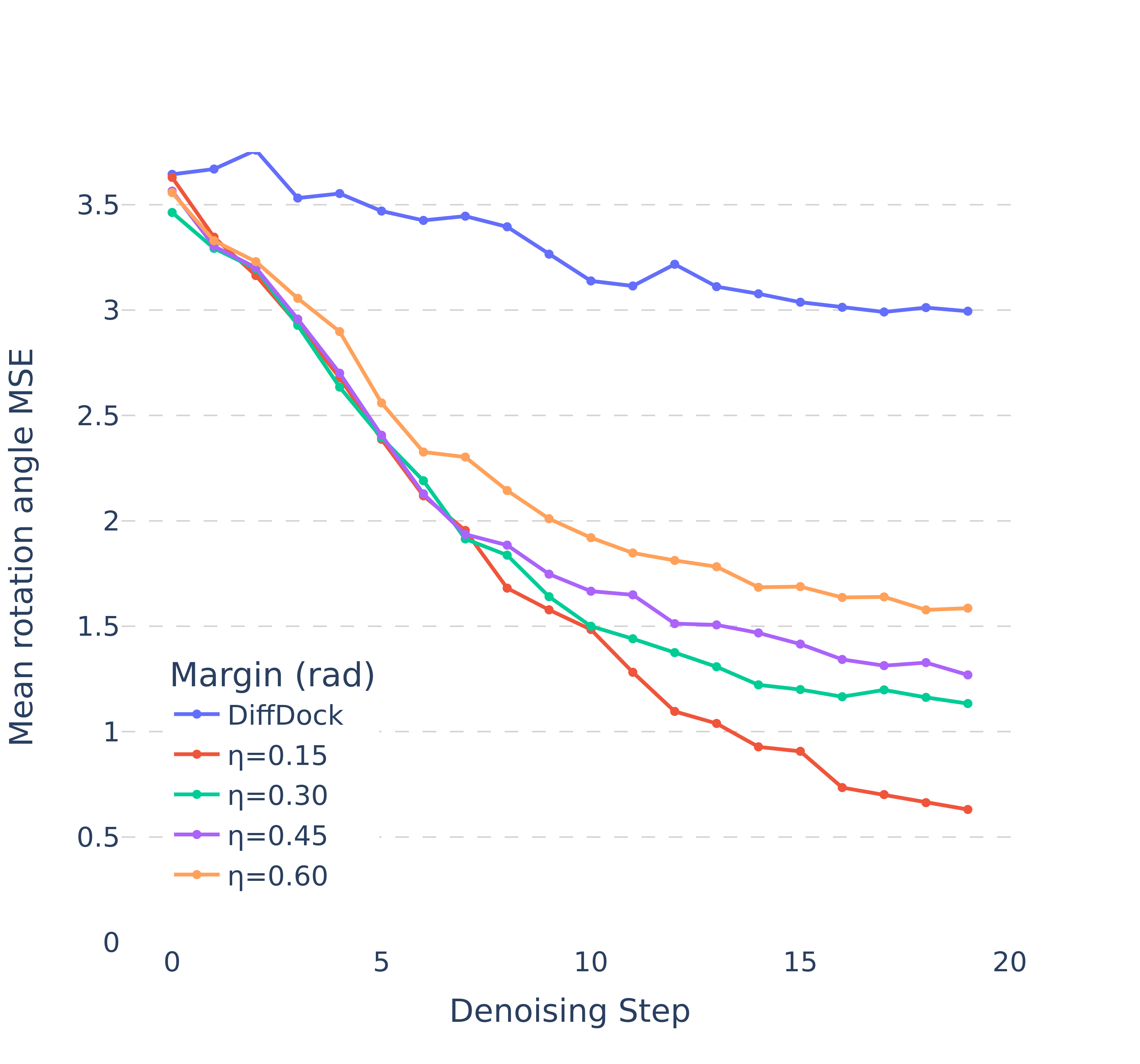}

  \label{fig:sub1}
\end{subfigure}%
\begin{subfigure}{.5\textwidth}
  \centering
  \includegraphics[width=\linewidth]{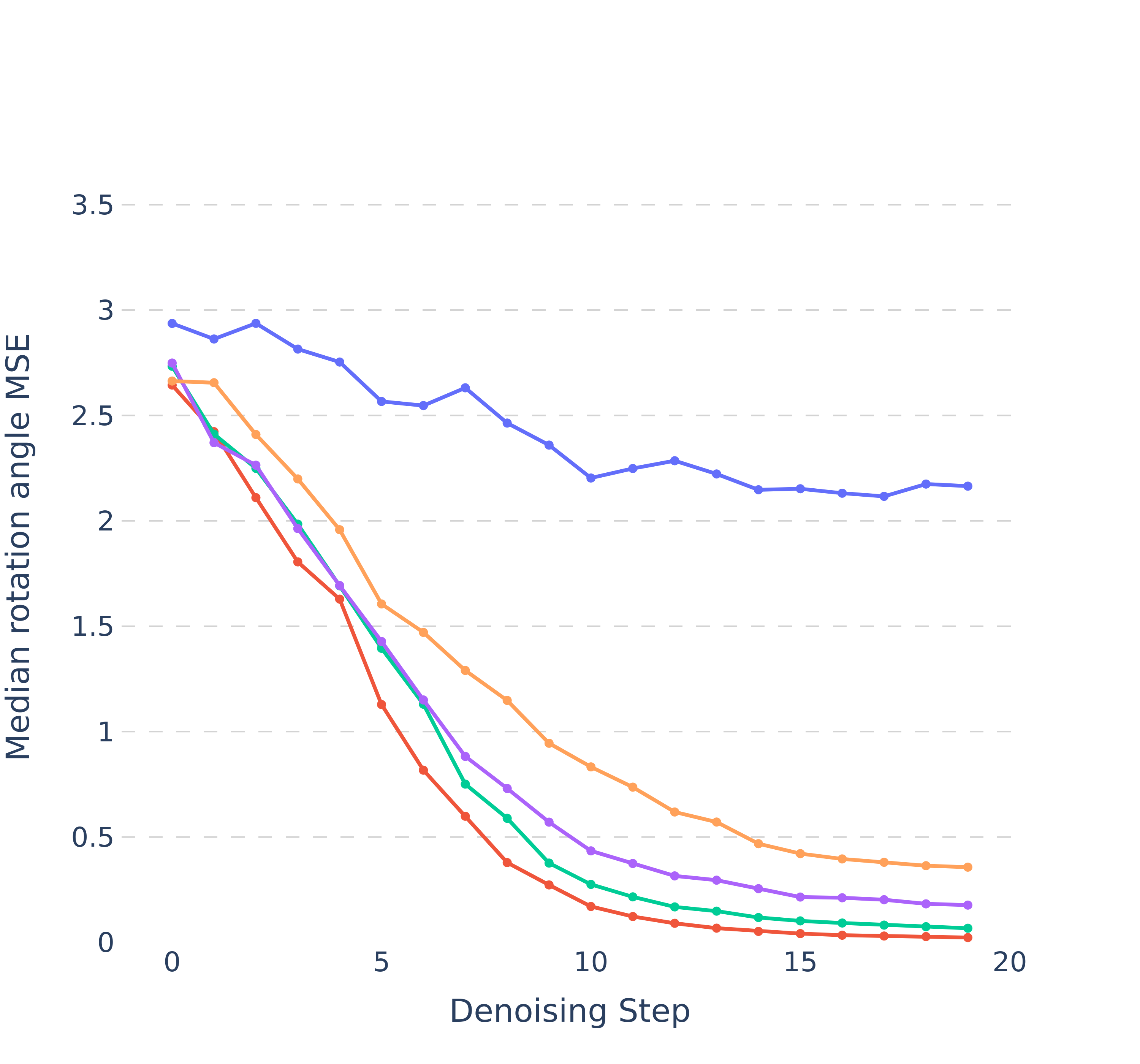}

  \label{fig:sub2}
\end{subfigure}
\caption{Mean and median MSE of rotation states versus denoising step.}
\label{fig:test}
\end{figure}

In this ablation we are able to determine an important reduction in rotation MSE, both in mean and median, with high robustness to the hyperparameter $\eta$ (measured in radians). 

We are able to see a clear degradation in performance with an increase in the region, however, even the worst-performing ablation has a much better performance than DiffDock in rotation MSE.

\subsection{Torsion}
In torsion guidance we also set the same default $\eta$ value for all the $\theta$
angles at 0.15. We evaluated values of 0.3, 0.45 and 0.6 for all the angles.

\begin{figure}[H]
\centering
\begin{subfigure}{.5\textwidth}
  \centering
  \includegraphics[width=\linewidth]{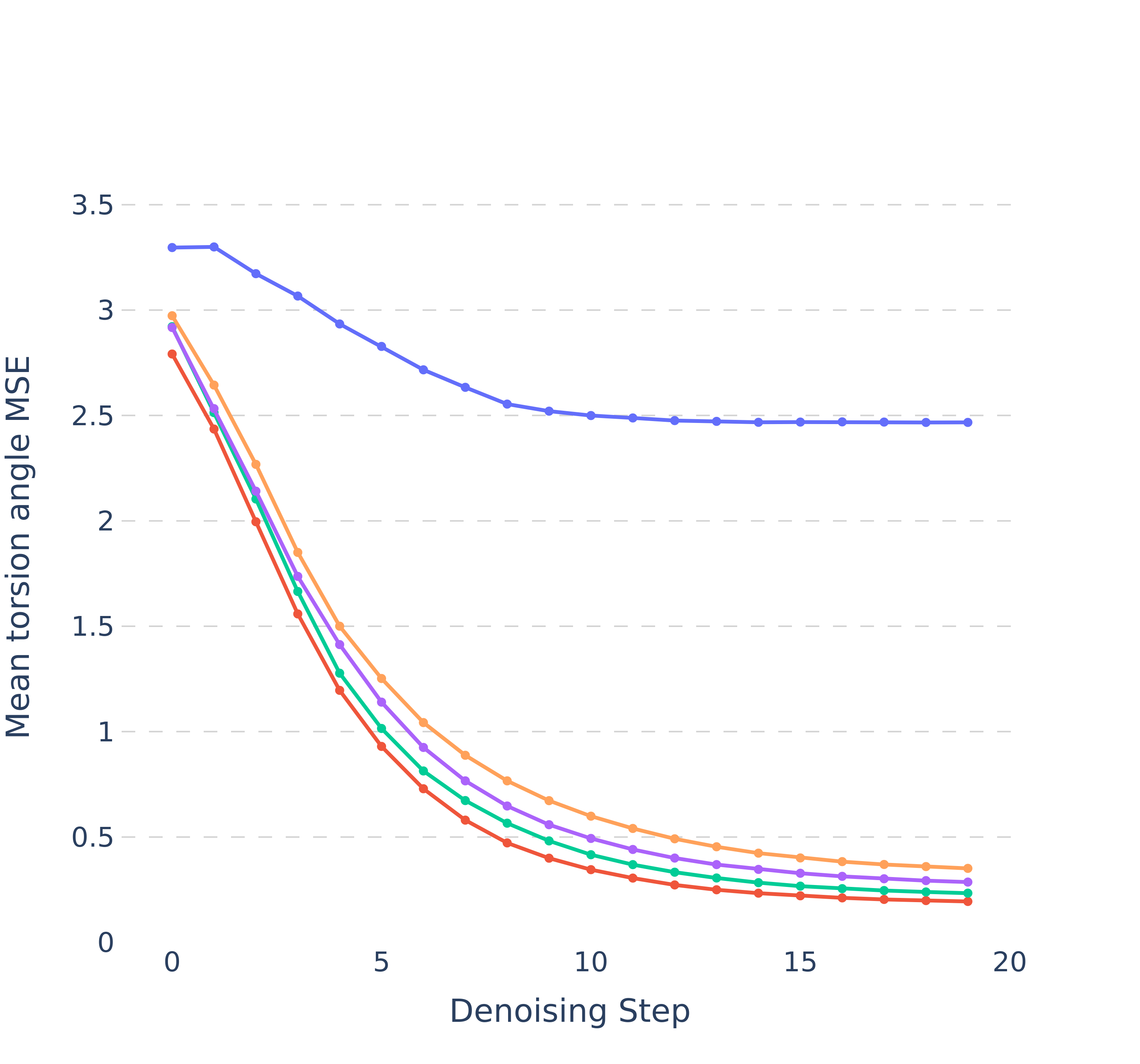}
  \label{fig:sub1}
\end{subfigure}%
\begin{subfigure}{.5\textwidth}
  \centering
  \includegraphics[width=\linewidth]{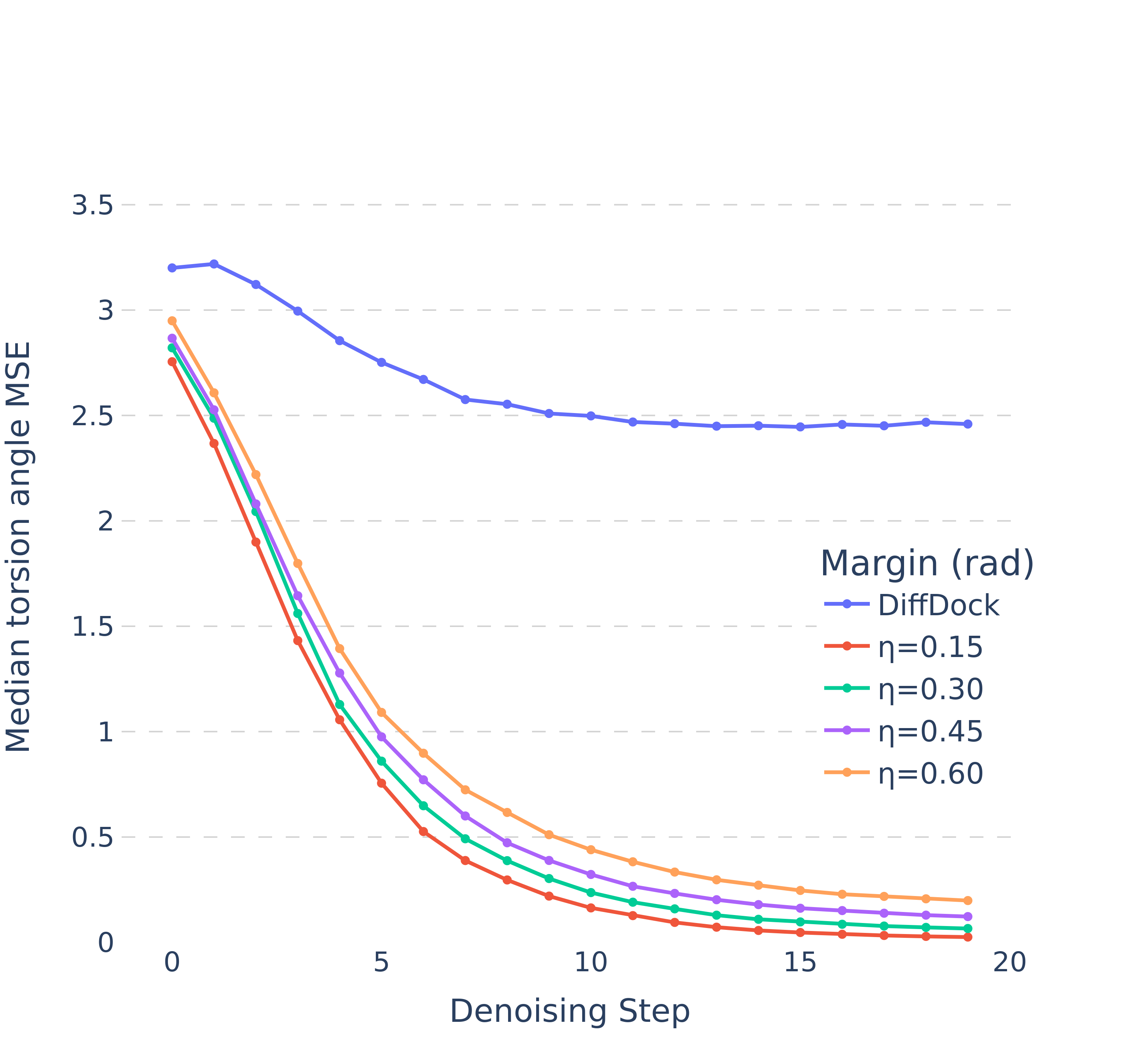}
  \label{fig:sub2}
\end{subfigure}
\caption{Mean and median MSE of torsion angles versus denoising step.}
\label{fig:test}
\end{figure}

This ablation follows the trend set in the rotation analysis, with a clear increase in performance for MSE even with higher values of $\eta$, alongside an accelerated convergence rate.

We are also able to determine this $\eta$ to be more robust than in rotation, with increasingly fuzzed regions performing very consistently with our base algorithm, and much better than DiffDock. This robustness analysis allowed us to begin extending this algorithm from self-docking to the MCS domain in \ref{exp:angletransf}.

\section{Guidance Strength Ablation} \label{gamma_ben}
In this section we evaluate the effect of altering the initial $\gamma$ hyperparameter, corresponding to the strength of the guidance that we impose upon the translation, rotation and torsion states. Lower values of gamma impose a weaker expert prior in the diffusion process and vice-versa. 

We test possible values in the range 0 to 0.5 with 0 being DiffDock and 0.2 being our default value for GDD. Following Appendix \ref{fuzz}, the comparison will be performed as the mean and median step-wise error of the 40 samples and 20 steps protocol, with gamma values of different subspaces varying at the same time.

\subsection{Translation}
\begin{figure}[H]
\centering
\begin{subfigure}{.5\textwidth}
  \centering
  \includegraphics[width=\linewidth]{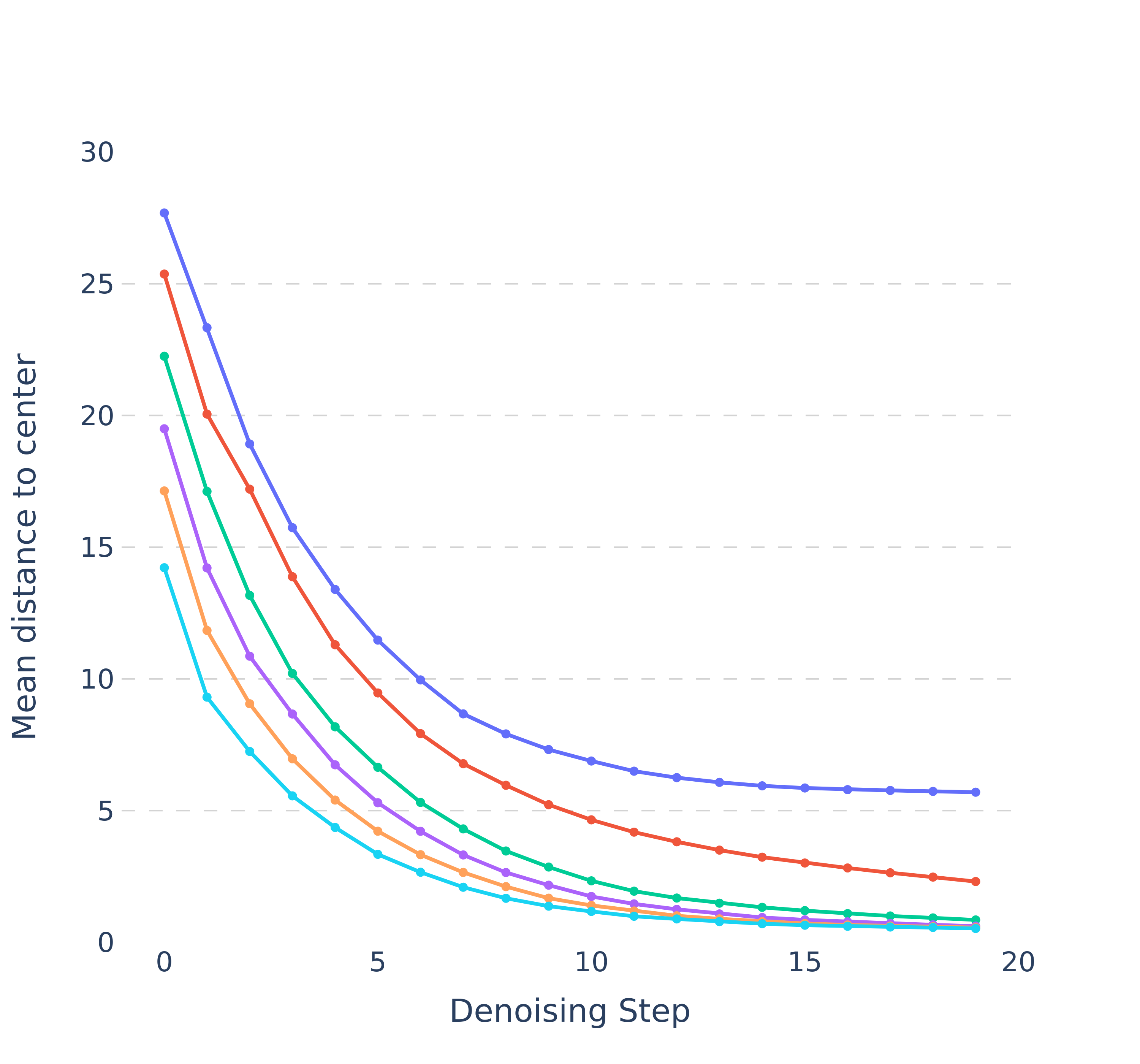}
  \label{fig:trsub1g}
\end{subfigure}%
\begin{subfigure}{.5\textwidth}
  \centering
  \includegraphics[width=\linewidth]{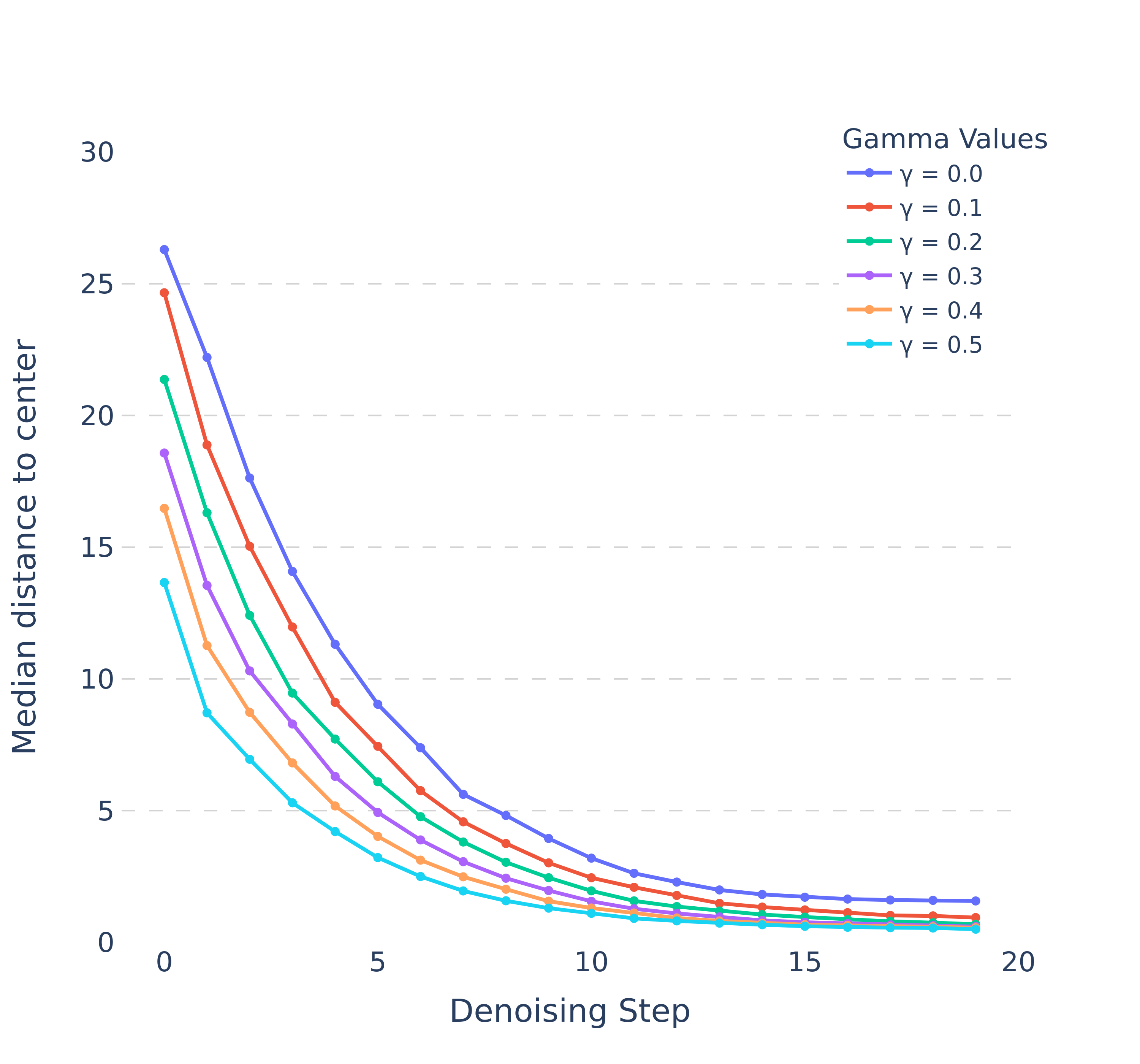}
  \label{fig:trsub2g}
\end{subfigure}
\caption{Mean and median distance to center of sphere versus denoising step.}
\label{fig:trg}
\end{figure}

As we are able to determine from Figure \ref{fig:trg}, an increase of the guidance strength allows for a faster convergence in steps compared to DiffDock. At the same time, a mild guidance of 0.1 already achieves the goal of reducing the amount of outliers in distance to the true center, as determined by the reduction in mean distance, with 0.2 achieving results very close to the ground truth. The rest of the range of values, however, do not significantly improve the previous result.

\subsection{Rotation}

\begin{figure}[H]
\centering
\begin{subfigure}{.5\textwidth}
  \centering
  \includegraphics[width=\linewidth]{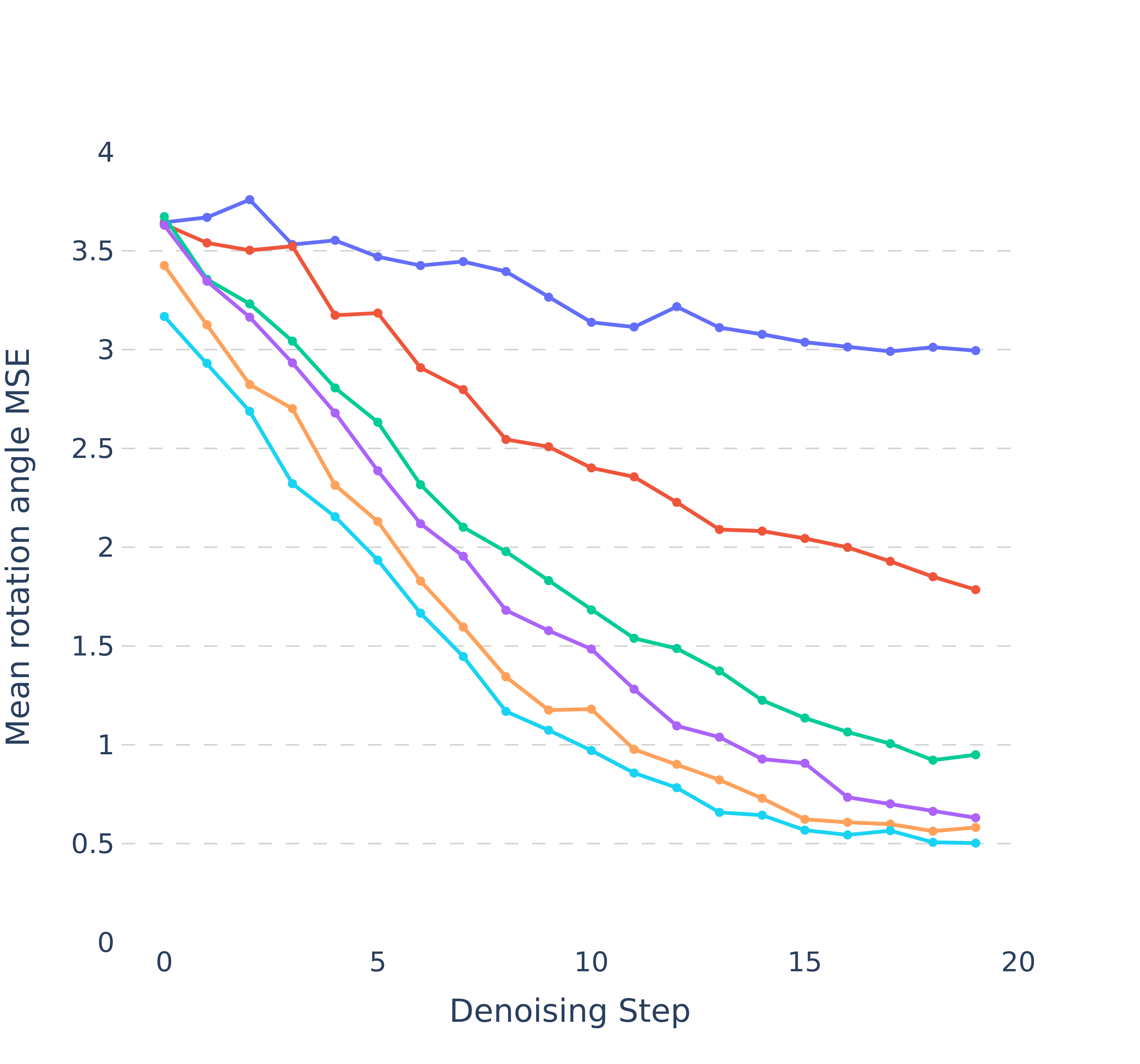}

  \label{fig:sub1gr}
\end{subfigure}%
\begin{subfigure}{.5\textwidth}
  \centering
  \includegraphics[width=\linewidth]{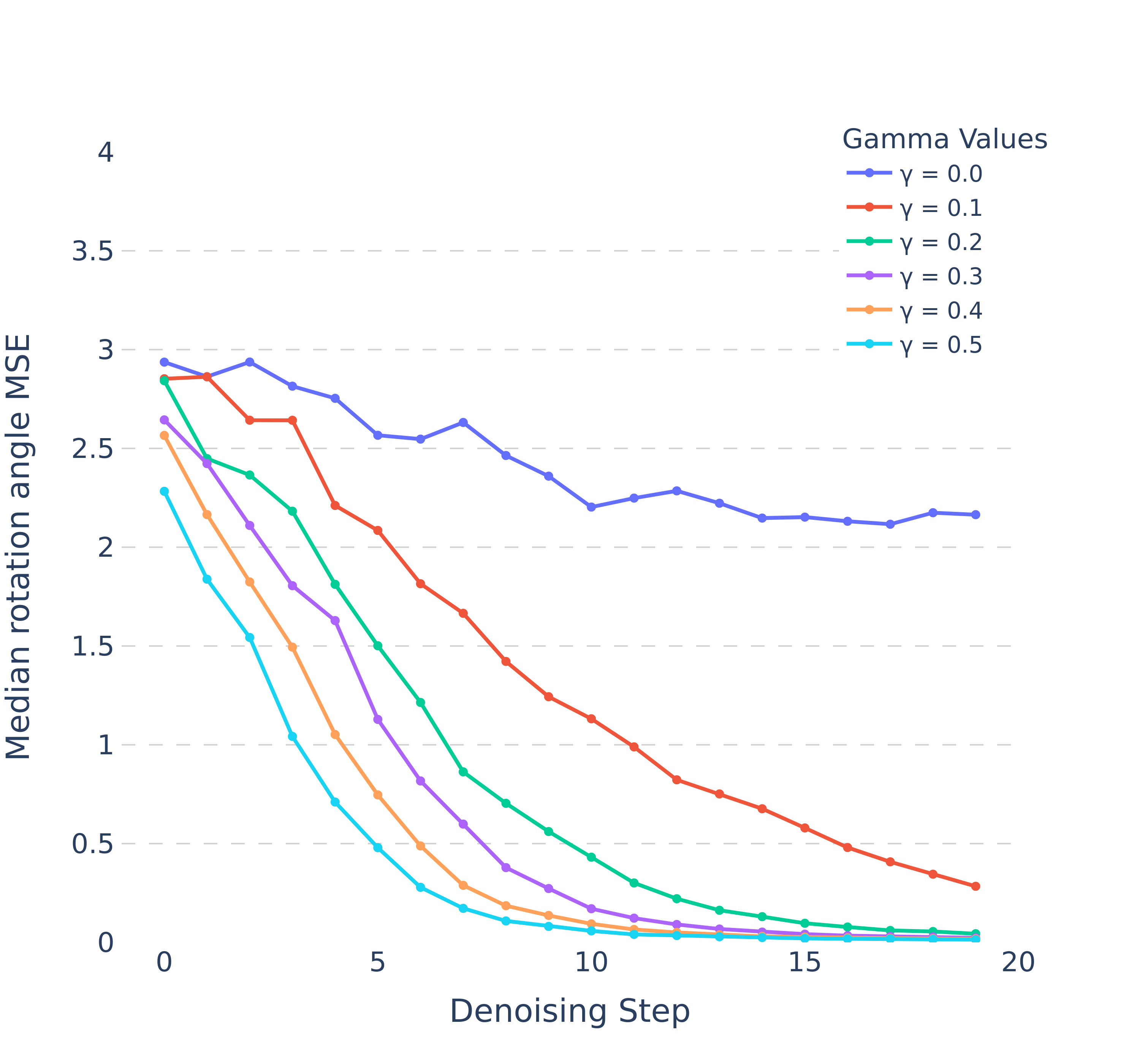}

  \label{fig:sub2gr}
\end{subfigure}
\caption{Mean and median MSE of rotation states versus denoising step.}
\label{fig:testgrot}
\end{figure}

In rotation, we are able to observe a very similar trend (Figure \ref{fig:testgrot}) to translation, with gamma values above 0.3 providing mostly faster convergence rather than an increase of performance, and with mild guidance values already providing an increase in performance compared to DiffDock.

\subsection{Torsion}

\begin{figure}[H]
\centering
\begin{subfigure}{.5\textwidth}
  \centering
  \includegraphics[width=\linewidth]{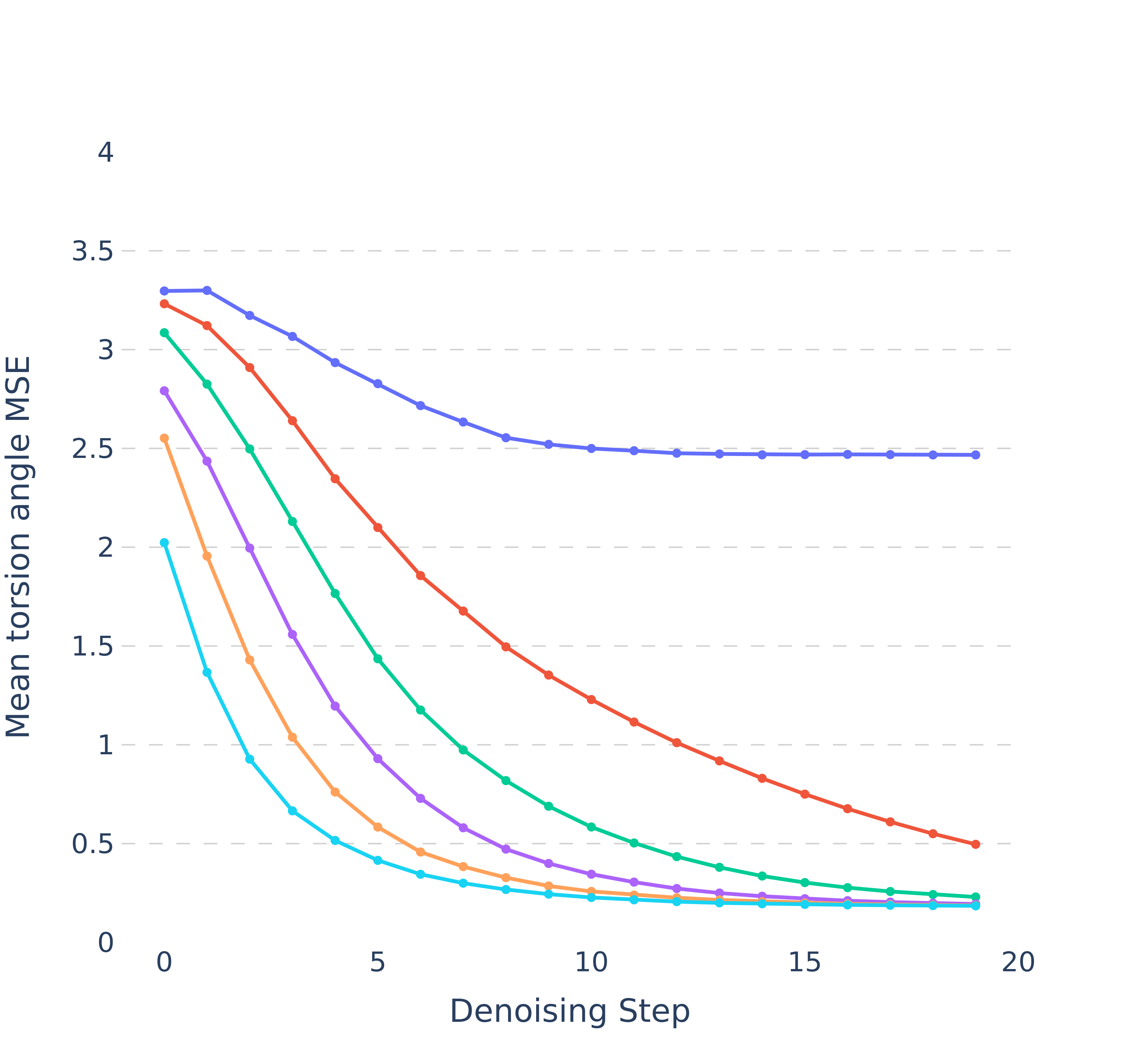}
  \label{fig:sub1gt}
\end{subfigure}%
\begin{subfigure}{.5\textwidth}
  \centering
  \includegraphics[width=\linewidth]{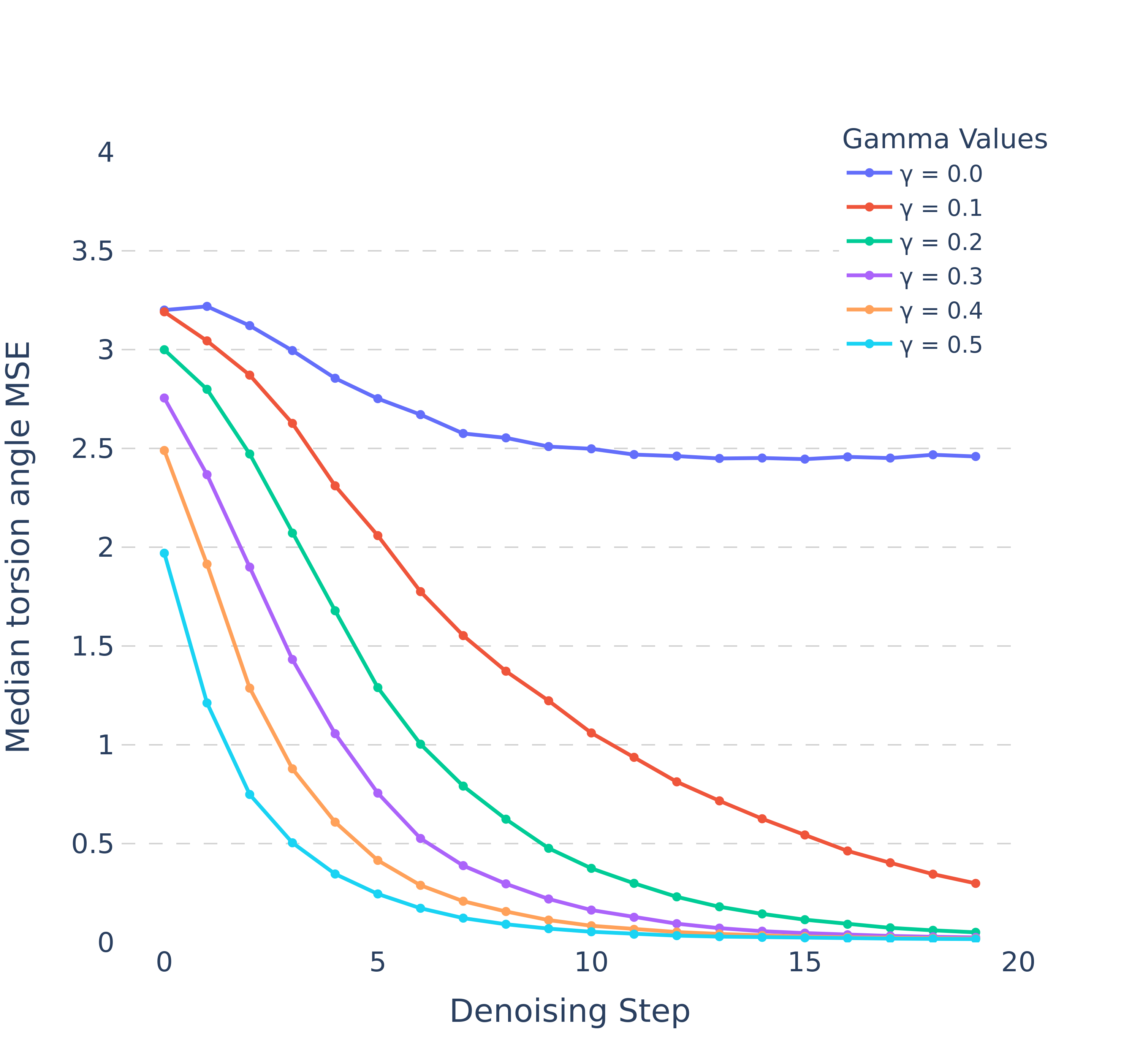}
  \label{fig:sub2gt}
\end{subfigure}
\caption{Mean and median MSE of torsion angles versus denoising step.}
\label{fig:testgt}
\end{figure}

Lastly, a very similar trend can be observed in torsion (Figure \ref{fig:testgt}) with values over 0.2 providing only convergence benefits, rather than performance increase. 

Overall, GeoDirDock maintains a robust performance across gamma values, with lower values of guidance already being enough for a increase of performance and higher values leading to equivalent performance and accelerated convergence.

\section{Generalization Across Protein Domains}\label{Dockgen_benchmark}
Generalization across different parts of the proteome is a desired property for deep learning docking methods that has been questioned in \citet{dockgen}. In this work, the authors develop a novel benchmark, DockGen, which contains a more chemically diverse set of binding pockets, compared to the ones contained in the original PDBBind test dataset.

\begin{table}[h]
\centering
\caption{RMSD and MSE performance comparison for the DockGen benchmark, showing Top-1 and Top-5 prediction accuracies. Best performances are highlighted in \textbf{bold}. Best performances with only translation guidance are stated in \textit{italic}. The number of denoising steps and the number of samples generated are stated as (steps-samples).}
\begin{tabularx}{\textwidth}{l|*{8}{>{\centering\arraybackslash}X}}
\hline
                & \multicolumn{2}{c|}{Top-1 RMSD}      & \multicolumn{2}{c|}{Top-5 RMSD}      & \multicolumn{2}{c|}{Top-1 MSE}      & \multicolumn{2}{c}{Top-5 MSE}      \\ 
                & \%$<$2       & Med      & \%$<$2       & Med      & Rot      & Tor     & Rot     & Tor     \\ \hline
DiffDock (10-10) & 2.27       & 9.48          & 7.39       & 5.82          & 3.70        & 2.56         & 1.13      & 1.39         \\
DiffDock (20-10) & 4.49       & 10.02          & 8.43       & 5.79          & 3.14       & 2.69         & 1.13      & 1.32         \\
DiffDock (20-40) & 3.93       & 9.61          & 8.99       & 6.08          & 3.20        & 2.69         & 1.02      & 1.42         \\ \hline
GDD-TR (10-10)   & 5.06       & 6.62          & 6.18       & 5.38          & 3.35        & 2.82         & 1.10      & 1.57         \\
GDD-TR (20-10)   & 4.52       & \textit{6.23}          & 8.47       & 4.82          & \textit{2.80}       &  \textit{2.66}       & 0.93      & 1.41         \\
GDD-TR (20-40)   & \textit{5.62}     &  6.29          &  \textit{8.99} &  \textit{4.63} &  2.83      & 2.72         &  \textit{0.89}      &  \textit{1.58}         \\ \hline
GDD-Full (10-10) & 10.11       & 5.81          & 13.48       & 4.91          & 0.83      & 0.61         & 0.12      & 0.28         \\
GDD-Full (20-10) & 16.29       & 4.91          & 20.22       & 4.27          & \textbf{0.17}     & \textbf{0.10}         & 0.05      & \textbf{0.06}         \\
GDD-Full (20-40) & \textbf{17.98} & \textbf{4.75} & \textbf{24.16}       & \textbf{3.87}          & 0.20 & \textbf{0.10} & \textbf{0.04} & \textbf{0.06} \\ \hline
\end{tabularx}
\label{tab:dockgen}
\end{table}

In these results we are able to determine an increase in performance of both GDD-TR and GDD-Full, following the trend of section \ref{exp:rmsd}. This highlights the capabilities of our method to generalize across a diverse set of binding modes.

\end{document}